\documentclass[aps,prl,twocolumn,showpacs,groupedaddress]{revtex4}  
\usepackage{graphicx}  
\usepackage{dcolumn}   
\usepackage{bm}        
\usepackage{amssymb}   

\begin{document}

\hspace{5.2in} \mbox{FERMILAB-PUB-08-089-E}

\title{Measurement of the polarization of the $\Upsilon(1S)$
and $\Upsilon(2S)$ states in $p\overline{p}$ collisions at $\sqrt{s}=$1.96\,TeV} 
%
\author{V.M.~Abazov$^{36}$}
\author{B.~Abbott$^{75}$}
\author{M.~Abolins$^{65}$}
\author{B.S.~Acharya$^{29}$}
\author{M.~Adams$^{51}$}
\author{T.~Adams$^{49}$}
\author{E.~Aguilo$^{6}$}
\author{S.H.~Ahn$^{31}$}
\author{M.~Ahsan$^{59}$}
\author{G.D.~Alexeev$^{36}$}
% ...

\affiliation{\vspace{0.1 in}(The D\O\ Collaboration)\vspace{0.1 in}}
% ...
\affiliation{$^{6}$University of Alberta, Edmonton, Alberta, Canada,
                Simon Fraser University, Burnaby, British Columbia, Canada,
                York University, Toronto, Ontario, Canada, and
                McGill University, Montreal, Quebec, Canada}
% ...
\affiliation{$^{29}$Tata Institute of Fundamental Research, Mumbai, India}
% ...
\affiliation{$^{31}$Korea Detector Laboratory, Korea University, Seoul, Korea}
% ...
\affiliation{$^{36}$Joint Institute for Nuclear Research, Dubna, Russia}
% ...
\affiliation{$^{49}$Florida State University, Tallahassee, Florida 32306, USA}
% ...
\affiliation{$^{51}$University of Illinois at Chicago,
                Chicago, Illinois 60607, USA}
% ...
\affiliation{$^{59}$Kansas State University, Manhattan, Kansas 66506, USA}
% ...
\affiliation{$^{65}$Michigan State University,
                East Lansing, Michigan 48824, USA}
% ...
\affiliation{$^{75}$University of Oklahoma, Norman, Oklahoma 73019, USA}
% ...
\date{October 31, 2008}

\begin{abstract}
We present a study of the polarization of the $\Upsilon(1S)$ and
$\Upsilon(2S)$ states using a $1.3\,$ fb$^{-1}$ data sample collected
by the D0 experiment in 2002--2006 during Run II of the Fermilab Tevatron Collider.
We measure the polarization parameter 
$\alpha=(\sigma_{T}-2\sigma_{L})/(\sigma_{T}+2\sigma_{L})$,
where $\sigma_{T}$ and $\sigma_{L}$ are the transversely and longitudinally
polarized components of the production cross section, as a function
of the transverse momentum ($p_{T}^{\Upsilon}$) for the $\Upsilon(1S)$ and $\Upsilon(2S)$.
Significant $p_{T}^{\Upsilon}$-dependent longitudinal polarization
is observed for the $\Upsilon(1S)$. A comparison with theoretical models is presented.
\end{abstract}

\maketitle 


The production of heavy quarks and quarkonium states at high energies
is under intense experimental and theoretical study~\cite{Brambila}.
The non-relativistic QCD (NRQCD) factorization approach has been developed
to describe the inclusive production and decay of quarkonia~\cite{cn1}
including high transverse momentum ($p_{T}$) $S$-wave charmonium
production at the Fermilab Tevatron Collider~\cite{cn2}. The theory introduces several
nonperturbative color-octet matrix elements (MEs). These MEs are universal
and are fitted to data of the Fermilab Tevatron Collider~\cite{fitME}. 
The universality of the MEs has
been tested in various experimental situations~\cite{cn3}. A remarkable
prediction of the NRQCD approach is that the $S$-wave quarkonium produced
in the $p\overline{p}$ collision should be transversely polarized
at sufficiently large $p_{T}$~\cite{cn4}. This prediction is based on
the dominance of gluon fragmentation in quarkonium production at large
$p_{T}$~\cite{cn2} and on the approximate heavy-quark spin symmetry
of NRQCD~\cite{cn1}. Measurements of the polarization of prompt $J/\psi$
by the CDF Collaboration do not confirm this prediction~\cite{cn54}.
 
A convenient measure of the polarization is the variable 
\begin{equation}
\alpha=(\sigma_{T}-2\sigma_{L})/(\sigma_{T}+2\sigma_{L}),
\end{equation}
where $\sigma_{T}$ and $\sigma_{L}$ are the transversely and longitudinally
polarized components of the production cross section. If we consider the decays
of quarkonium to a charged lepton-antilepton pair, then the angular distribution
is given by
\begin{equation}
\frac{dN}{d(\cos\theta^{\ast})} \propto 1+\alpha\cos^{2}\theta^{\ast},
\end{equation}
where $\theta^{\ast}$ is the angle of the positive lepton in the quarkonium center-of-mass
frame with respect to the momentum of the decaying particle in the
laboratory  frame.

Quantitative calculations of the polarization for inclusive $\Upsilon(nS)$
mesons are carried out~\cite{cn6} by using the ME for direct bottomonium
production determined from an analysis of Tevatron data~\cite{cn7}.
They predict that  the transverse polarization
of $\Upsilon(1S)$ should dominate and increase steadily with $p_{T}^\Upsilon$ 
for $p_{T}^{\Upsilon} \gtrsim
10$ GeV/$c$ and that the $\Upsilon(2S)$ and $\Upsilon(3S)$
should be even more strongly transversely polarized. The $k_t$-factorization
model~\cite{cn8}, using a semi-hard approach, predicts a longitudinal
polarization of $\Upsilon(1S)$ at $p_{T}^{\Upsilon}>5\,$GeV/$c$~\cite{kt}. 
In this context, the experimental measurement of the $\Upsilon$
polarization is a crucial test of two theoretical approaches to parton
dynamics in QCD.

The D0 detector is described in detail elsewhere~\cite{d0det}.  The main
elements relevent to this analysis are a central-tracking system, consisting of a 
silicon microstrip tracker (SMT) and a central fiber tracker (CFT), 
and muon detector systems.

The data set used for this analysis includes approximately 1.3 fb$^{-1}$
of integrated luminosity collected by the D0 detector between April
2002 and the end of 2006. 
We selected events where the $\Upsilon(nS)$ decayed into two muons. Muons 
were required to have hits in three muon layers, to have an associated
track in the central tracking system with hits in both the SMT and
CFT, and to have transverse momentum $p_{T}^{\mu}>3.5$\,GeV/$c$. In this analysis
only events that passed a dimuon trigger, which requires two opposite charge muon candidates,
  were included in the final sample. 
We observed about 260,000 $\Upsilon(nS)$ with rapidity 
$\mid\! y^{\Upsilon}\!\mid<1.8$ 
when fitting the  dimuon invariant mass distribution as described below. 

Monte Carlo (MC) samples for unpolarized $\Upsilon(1S)$ and 
$\Upsilon(2S)$ inclusive production 
were generated using the {\sc pythia}~\cite{pythia} event
generator and then passed through a {\sc geant}-based~\cite{geant} simulation
of the D0 detector. The simulated
events were then required to satisfy the same selection criteria as the
data sample including a detailed simulation of all aspects of the trigger 
requirements.
 
We fitted the dimuon invariant mass distribution in several intervals
of $p_{T}^{\Upsilon}$ for a set of $|\cos\theta^{\ast}|$ bins. A previous measurement
of the $\Upsilon(1S)$ cross-section by the D0 experiment~\cite{upsx} showed
that a double Gaussian function is required to model the mass distribution
of the $\Upsilon(1S)$ candidates. 
Studies performed on the $\Upsilon(1S)$ Monte Carlo sample suggest that a more
sophisticated parameterization of the invariant mass distribution for
some  $|\cos\theta^{\ast}|$ bins, where we observe non-Gaussian tails, is required. 
Two different parameterizations of the mass distribution
were used, referred to as ``data-driven'' and ``MC-driven'' functions.
The data-driven function has the advantage that no assumptions are made about 
how well the MC reproduces the true resolution. It consists of a double Gaussian
function with equal means. The mean, widths, and relative fraction are free parameters.
In contrast, the MC-driven function allows for a test of the effect of non-Gaussian components 
to the resolution that are observable in MC but are hidden in data 
by the detector resolution and the combinatoric background. Non-Gaussian tails are 
implemented via a third Gaussian component with a 
floating mean to account for an asymmetric tail in the 
reconstructed $\Upsilon(nS)$ mass.  The width and relative fraction are 
taken from Monte Carlo.
\begin{figure}
\includegraphics[scale=1.]{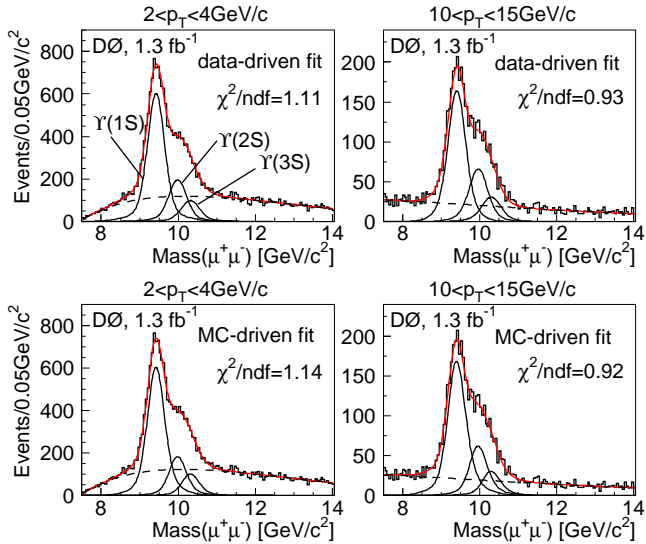}
\caption{\label{fig:data_ext} [Color online] Signal extraction from the dimuon invariant mass 
           distribution for events in the $0.4 < |\cos\theta^{\ast}| < 0.5$ region. 
	   a,c) $2<p_{T}^{\Upsilon}<4$\,GeV/$c$;  
	   b,d) $10<p_{T}^{\Upsilon}<15$\,GeV/$c$. Dashed curves are
	   the combinatoric background.}
\end{figure}
Figure~\ref{fig:data_ext} shows an example of a fit to the mass distribution for
a single $p_T^{\Upsilon}$ and $|\cos\theta^{\ast}|$ bin ignoring or including 
non-Gaussian tails. The signal consists of three
mass peaks, the $\Upsilon(1S)$, $\Upsilon(2S)$, and $\Upsilon(3S)$
where the mass differences were fixed to the measured values~\cite{pdg}.
The background was modeled with a convolution of an exponential and
a polynomial function. The degree of the polynomial was chosen to be between
one and six depending on the complexity of the shape of the background. 
The $\chi^2$ values  in Fig.~\ref{fig:data_ext} do not allow us to 
differentiate between the two approximations and hence we average them.

The data were divided into bins in $p_{T}^{\Upsilon}$ and
$|\cos\theta^{\ast}|$. For each of these bins the numbers of $\Upsilon(1S)$
and $\Upsilon(2S)$ candidates were extracted from the mass distribution.
The number of $\Upsilon(3S)$ candidates was insufficient
to extract angular distributions. 

\begin{figure}
\includegraphics[scale=1.]{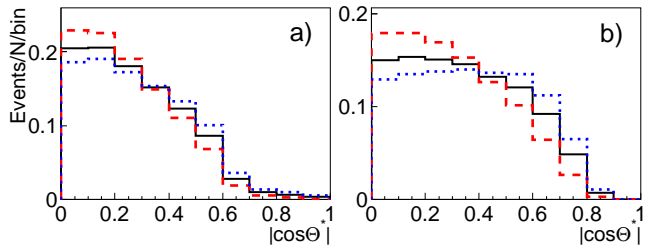}
\vspace{-.2cm}
\caption{\label{fig:sensivity_mc} [Color online] Monte Carlo 
  $|\cos\theta^{\ast}|$ distributions after all selection requirements
  for different $\alpha$ values: $-1$ (dashed histogram), $0$ 
  (solid histogram) and $+1$
  (dotted histogram). a) $0<p_{T}^{\Upsilon}<1$\,GeV/$c$, 
  b) $p_{T}^{\Upsilon}>$15\,GeV/$c.$}
\end{figure}

\begin{figure}
\includegraphics[scale=1.]{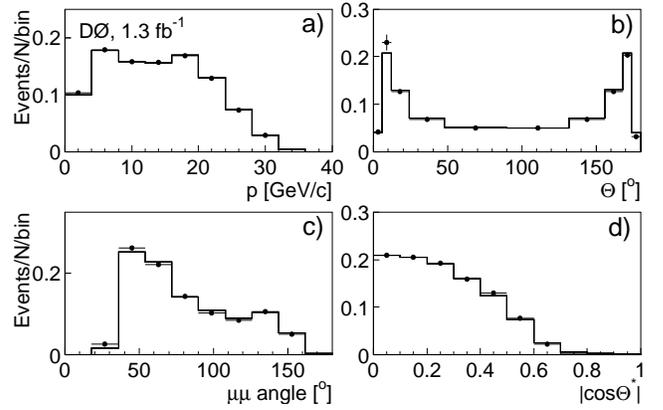}
\vspace{-.2cm}
\caption{\label{fig:data_mc} Comparison of data (points) and Monte Carlo 
  (solid histogram) for $\Upsilon(1S)$ with  
  $2<p_{T}^{\Upsilon}<4$\,GeV/$c$: a) momentum of $\Upsilon(1S)$, 
  b) polar angle of $\Upsilon(1S)$,
  c) angle between muons,
  d) $|\cos\theta^{\ast}|$.}
\end{figure}

Polarization was not taken into account in the Monte Carlo 
generation. To compare them
with data we calculated for each event the weight $w_{\alpha}$, which
converts the initial Monte Carlo $|\cos\theta^{\ast}|$ distribution
with $\alpha=0$ to a distribution with the chosen $\alpha$. 
Figure~\ref{fig:sensivity_mc} shows 
the sensivity of the D0 detector to the $\Upsilon(1S)$ polarization for the
lowest and highest $p_{T}^{\Upsilon(nS)}$ intervals.
The {\sc pythia} simulation does not accurately model the kinematic
distributions of $\Upsilon(nS)$  production at the Tevatron 
(e.g., the $p_{T}^{\Upsilon(nS)}$
distribution). To correct the Monte Carlo distributions, we introduced additional weights 
to improve the agreement with data of the $\Upsilon(nS)$ momentum distribution. 
Instead of the weight $w_{\alpha}$ in our algorithm, we used the weight 
$w=w_{\alpha}w_{p_{^{T}}^{\Upsilon}}w_{p^{\Upsilon}}$,
where $w_{p_{^{T}}^{\Upsilon}}$ and $w_{p^{\Upsilon}}$ are weights
to achieve agreement between data and Monte Carlo distributions of
$p_{T}^{\Upsilon}$ and $p^{\Upsilon}$. 
After this reweighting procedure, we obtained good agreement between
data and MC for the $\Upsilon(nS)$ and muon kinematic distributions.
An example for Y(1S) with $2<p_{T}^{\Upsilon}<4$\,GeV/$c$, using the MC-driven fit, is
presented in Fig~\ref{fig:data_mc}.  
All data distributions were derived by
estimating the number of $\Upsilon(1S)$ events from a fit to the
dimuon mass distribution for the corresponding  
bin of the histogram. 

The systematic uncertainties on $\alpha$ for $\Upsilon(1S)$  are summarized
in Table~\ref{tab:table1}. Values of $\alpha$ were found for several $p_{T}^{\Upsilon}$ intervals, 
using both parameterizations (data-driven and MC-driven) of the 
dimuon invariant mass distribution for the signal. Both $\alpha$ measurements 
are averaged and one half of the difference between them is assigned
as systematic uncertainty due to the signal model. The uncertainty 
in the background was estimated by varying the mass range of the fit 
and the degree of the polynomial used to parameterize the background.
The MC simulation does not reproduce exactly the mass of the $\Upsilon(1S)$
peak, which differs by about 40\,MeV/$c^2$ from the PDG value. 
The effect 
on the $\alpha$ determination was estimated and shown in Table~\ref{tab:table1} under
``muon momentum.'' Finally, the systematic uncertainty due to the 
trigger simulation has also been considered and shown in Table~\ref{tab:table1}.
The $\Upsilon(1S)$ polarization was calculated assuming that it is constant within
a given $p_{T}^{\Upsilon}$ bin. This assumption leads to a small bias in the measured $\alpha$
that is estimated by reweighting the simulation using the observed
$p_{T}^{\Upsilon}$ dependence of $\alpha$. The final measured $\alpha$ is corrected by a 
factor ranging between $-$0.03 and $+$0.06, depending on $p_{T}^{\Upsilon}$.
\begin{table}
\caption{\label{tab:table1}Systematic uncertainties  on $\alpha$ for $\Upsilon(1S)$.}
\begin{ruledtabular}
\begin{tabular}{c|c|c}
Source & Uncertainty on $\alpha$\footnotemark[1] &
$p_{T}^{\Upsilon}$ 
\footnotemark[2] [GeV/$c$]\\
\hline
 Signal model       & $0.01-0.15$ & $1-2$ \\
 Background model   & $0.04-0.21$ & $0-1$ \\
 Muon momentum      & $0.00-0.06$ & $0-1$ \\
 Trigger simulation & $0.00-0.06$ & $>$15 \\
\end{tabular}
\end{ruledtabular}
\footnotetext[1]{For all $p_{T}^{\Upsilon}$ intervals}
\footnotetext[2]{Interval with maximal uncertainty}
\end{table}

\begin{figure}
\includegraphics[scale=0.33]{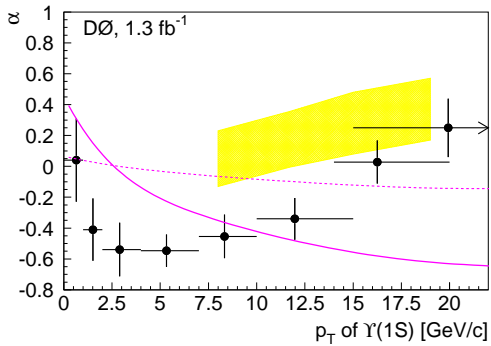}
\vspace{-.2cm}
\caption{\label{fig:data_u1s} [Color online] Dependence of $\alpha$ on $p_{T}^{\Upsilon}$ 
for inclusive $\Upsilon(1S)$ candidates.
Black circles are data. The band is the NRQCD prediction~\cite{cn6}.
Curves are two limiting cases (see text) of the $k_t$-factorization
model ~\cite{kt}.}
\end{figure}

\begin{figure}
\includegraphics[scale=0.33]{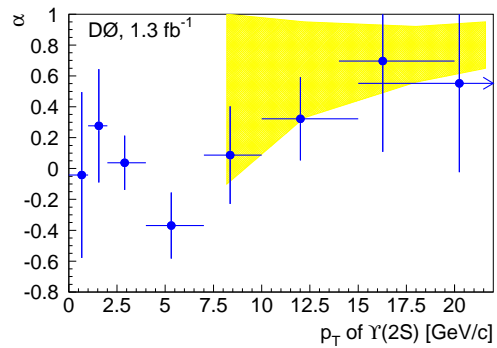}
\vspace{-.2cm}
\caption{\label{fig:data_u2s} [Color online] Dependence of $\alpha$ on $p_{T}^{\Upsilon}$ 
for inclusive $\Upsilon(2S)$ production.
Circles are our data. The band is the NRQCD prediction~\cite{cn6}.}
\end{figure}

Figure~\ref{fig:data_u1s} shows the measured $\alpha$ as a function of 
$p_{T}^{\Upsilon}$ for $\Upsilon(1S)$. 
Note that the bin for 14-20 GeV is not statistically independent from
the adjacent bins. 
The arrow indicates that the highest $p_T^{\Upsilon}$ interval considered,
$p_T^{\Upsilon}>15$\,GeV/$c$, does not
have an upper limit. 
The uncertainties are the systematic and
statistical uncertainties added in quadrature. Also shown are the
NRQCD prediction~\cite{cn6} (yellow band), and the two limits of the
$k_t$-factorization model~\cite{kt} (curves). The lower line corresponds
to the quark-spin conservation hypothesis, and the upper one to
the full quark-spin depolarization hypothesis. The previous measurement by CDF 
of the polarization of $\Upsilon(1S)$ with rapidity 
$\mid\! y^{\Upsilon}\!\mid<0.4$ is consistent with $\alpha$ equal to 
zero~\cite{cdf-upspol}.
We expect the CDF and D0 results 
to be similar and we have no explanation 
for the observed difference. We also extracted the polarization of  the 
$\Upsilon(2S)$, which is shown in Fig.~\ref{fig:data_u2s} along with the NRQCD
predictions~\cite{cn6}. Values of $\alpha$ for statistically independent 
$p_{T}^{\Upsilon}$ intervals, shown in 
Fig.~\ref{fig:data_u1s} and Fig.~\ref{fig:data_u2s}, are given in Table~\ref{tab:table2}.
\begin{table}
\caption{\label{tab:table2}Measurements of $\alpha$ for $\Upsilon(1S)$ and $\Upsilon(1S)$.}
\begin{ruledtabular}
\begin{tabular}{ccc}
$p_{T}^{\Upsilon}$ [GeV/$c$] & $\alpha[\Upsilon(1S)]$ &
$\alpha[\Upsilon(2S)]$ \\
\hline
  $0-1$       & $ 0.04\pm0.27$ & $-0.04\pm0.54$ \\
  $1-2$       & $-0.41\pm0.20$ & $ 0.28\pm0.37$ \\
  $2-4$       & $-0.54\pm0.17$ & $ 0.04\pm0.18$ \\
  $4-7$       & $-0.55\pm0.10$ & $-0.37\pm0.21$ \\
  $7-10$      & $-0.45\pm0.14$ & $ 0.09\pm0.32$ \\
  $10-15$     & $-0.34\pm0.14$ & $ 0.32\pm0.27$ \\
  $>$15       & $ 0.25\pm0.19$ & $ 0.55\pm0.58$ \\
\end{tabular}
\end{ruledtabular}
\end{table}

In conclusion, we have presented measurements of the polarization of the $\Upsilon(1S)$
and $\Upsilon(2S)$ as functions of $p_{T}^{\Upsilon}$ from 0\,GeV/$c$ to 20\,GeV/$c$.
Significant $p_{T}$-dependent longitudinal polarization
is observed for the $\Upsilon(1S)$ inconsistent with NRQCD
predictions. At $p_{T}^{\Upsilon}>$7\,GeV/$c$ the fraction of transversely 
polarized $\Upsilon(2S)$ particles is higher than in $\Upsilon(1S)$ at the same value of 
$p_{T}^{\Upsilon}$, in agreement with NRQCD predictions.

%
We thank the staffs at Fermilab and collaborating institutions, 
and acknowledge support from the 
DOE and NSF (USA);
CEA and CNRS/IN2P3 (France);
FASI, Rosatom and RFBR (Russia);
CNPq, FAPERJ, FAPESP and FUNDUNESP (Brazil);
DAE and DST (India);
Colciencias (Colombia);
CONACyT (Mexico);
KRF and KOSEF (Korea);
CONICET and UBACyT (Argentina);
FOM (The Netherlands);
STFC (United Kingdom);
MSMT and GACR (Czech Republic);
CRC Program, CFI, NSERC and WestGrid Project (Canada);
BMBF and DFG (Germany);
SFI (Ireland);
The Swedish Research Council (Sweden);
CAS and CNSF (China);
and the
Alexander von Humboldt Foundation.
%

\end{document}